\documentstyle[psfig,draft]{mn}

\def\simgt{\mathrel{\raise0.35ex\hbox{$\scriptstyle >$}\kern-0.6em
\lower0.40ex\hbox{{$\scriptstyle \sim$}}}}
\def\magcir{\mathrel{\raise0.35ex\hbox{$\scriptstyle >$}\kern-0.6em
\lower0.40ex\hbox{{$\scriptstyle \sim$}}}}
\def\simlt{\mathrel{\raise0.35ex\hbox{$\scriptstyle <$}\kern-0.6em
\lower0.40ex\hbox{{$\scriptstyle \sim$}}}}
\def\mincir{\mathrel{\raise0.35ex\hbox{$\scriptstyle <$}\kern-0.6em
\lower0.40ex\hbox{{$\scriptstyle \sim$}}}}

\newcommand{\be}{\begin{equation}}
\newcommand{\ee}{\end{equation}}
\newcommand{\ba}{\begin{eqnarray}}
\newcommand{\ea}{\end{eqnarray}}
\newcommand{\brr}{\begin{array}}
\newcommand{\err}{\end{array}}

\newcommand{\si}{\sigma}
\newcommand{\Dc}{\Delta_{\rm c}}
\newcommand{\omm}{\Omega_{\rm m}}
\newcommand{\oml}{\Omega_{\Lambda}}

\input epsf

\begin{document}

\title[Cluster abundance normalization]{On determining the cluster abundance
normalization}

\author[E. Pierpaoli, S. Borgani, D. Scott \& M. White] 
{Elena Pierpaoli${}^1$, Stefano Borgani${}^2$, Douglas Scott${}^3$
 and Martin White${}^4$\\
${}^1$Physics and Astronomy Department, Princeton University, Princeton,
NJ, 08544~~USA\\
${}^2$Dipartimento di Astronomia, Universit\`a di Trieste, via Tiepolo
11, I-34131 Trieste~~Italy\\
${}^3$Department of Physics and Astronomy, University of British Columbia,
Vancouver, B.C., V6T 1Z1~~Canada\\
${}^4$Departments of Physics and Astronomy, University of California,
Berkeley, CA, 94720~~USA\\}

\date{Accepted ... ;
      Received ... ;
      in original form ...}

\pagerange{000--000}

\maketitle

\begin{abstract}
Different determinations currently suggest scattered values for the
power spectrum normalization on the scale of galaxy clusters, $\si_8$.
Here we concentrate on the constraints coming from the X-ray
temperature and luminosity functions (XTF and XLF), and investigate
several possible sources of discrepancies in the results.
We conclude that the main source of error in both methods
is the scaling relation involved, in particular the way its intrinsic
scatter and systematic normalization are treated.

For temperature derived constraints, we use a sample adapted from
HIFLUGCS, and test for several sources of systematic error.  We
parameterize the mass-temperature relation with an overall factor
$T_\ast$, which varies between about 1.5 and 1.9 in the literature,
with simulations typically giving lower results than empirically
derived estimates.  After marginalising over this range of $T_\ast$,
we obtain a 68 per cent confidence range of
$\sigma_8=0.77^{+0.05}_{-0.04}$ for a standard
$\Lambda$CDM model.

For luminosity derived constraints we use the XLF from the REFLEX
survey and explore how sensitive the final results are on the details
of the mass--luminosity, $M$--$L$, conversion. Assuming a uniform
systematic uncertainty of $\pm20$ per cent in the amplitude of the
mass-luminosity relation by Reiprich \& B\"ohringer, we derive
$\sigma_8=0.79^{+0.06}_{-0.07}$ for the same standard $\Lambda$CDM
model. Although the XTF and XLF derived constraints agree very well
with each other, we emphasize that such results can change by about
10--15 per cent, depending on how uncertainties in the $L$--$T$--$M$
conversions are interpreted and included in the analysis.

We point out that in order to achieve precision cosmology on
$\sigma_8$ using cluster abundance, it is first important to separate
the uncertainty in the scaling relation into its intrinsic and overall
normalization parts.  Careful consideration of all sources of scatter
is also important, as is the use of the most accurate formulae and full
consideration of dependence on cosmology.  A significant improvement will
require the simultaneous determination of mass using a variety of distinct
methods, such as X-ray observations, weak lensing, Sunyaev-Zel'dovich
measurements and velocity
dispersions of member galaxies, for a moderately large sample of clusters.
\end{abstract}

\begin{keywords}
gravitation -- galaxies: clusters: general --
cosmology: theory -- large-scale structure of Universe
\end{keywords}

\section{Introduction}

The normalization of density perturbations on large scales is a
fundamental parameter describing our Universe.  Its determination has
been actively pursued for the last quarter century, using a wide range
of methods.  Now we are entering the era of precision cosmology, with
several cosmological parameters apparently determined to better than
10 per cent accuracy.  While this level of precision has been reached for the
amplitude of the largest scale perturbations, through the {\sl COBE\/}
anisotropy measurements, the direct normalization of matter
fluctuations on galaxy cluster scales still yields a wide range of
estimates.

Here we focus on the determination of the cluster mass distribution
via X-ray observations, and discuss how differences in the details of
the adopted strategies can lead to different, sometimes quite
discrepant, determinations of $\sigma_8$. While considering X-ray
observations, two approaches have traditionally been adopted. The
first one is based on determining the cluster X-ray luminosity
function (XLF) and then converting it into the mass function through a
suitable Mass--Luminosity, $M$--$L$, conversion (e.g.~Henry et
al. 1992; Sadat, Blanchard \& Oukbir 1998; Reichart et al.~1999;
Borgani et al.~2001; Reiprich \& B\"ohringer 2002; Viana, Nichol \&
Liddle 2002; Schuecker et al.~2002). The second possibility is to use
measurements of the X-ray temperature function (XTF) and then apply a
Mass--Temperature, $M$--$T$, relation (e.g.~Oukbir \& Blanchard 1992;
Eke et al.~1998; Viana \& Liddle 1998; Pen 1998; Markevitch 1998;
Pierpaoli, Scott \& White 2001; Seljak 2002).

Measuring cluster luminosities requires ${\sim}\,10$ times less X-ray
photons than measuring temperature. This has allowed us to precisely
determine the XLF for samples containing a few hundreds of local
clusters.  Current independent determinations of the XLF agree with
each other at high precision (see Rosati, Borgani \& Norman 2002, for
a review), thus making it a stable reference point to quantify the
cluster population.  However,the dependence of the bremsstrahlung
emissivity on the square of the local gas density causes the X-ray
luminosity to be highly sensitive to the core and to local details of
the intra-cluster medium (ICM) structure.  This makes it a noisy and
difficult to calibrate mass estimator.  Resorting to an
observationally calibrated $M$--$L$ relation is in principle possible
through two alternative routes: either combining the $L$--$T$ and the
$M$--$T$ relations, or directly measuring an $M$--$L$ scaling.  Both
approaches require an independent measure of $M$, which is a
notoriously difficult problem.  Traditionally, masses are estimated
either by applying hydrostatic equilibrium to a cluster with a
measured temperature pattern and surface brightness profile (Reiprich
\& B\"ohringer 2002; Ettori, De Grandi \& Molendi 2002), or using the
mass estimated through cosmic shear (e.g.~Viana \& Liddle 2002, Allen
et al.~2002).  Both of these methods could suffer biases (Evrard,
Metzler \& Navarro~1996; Metzler, White \& Loken~2001) which need to
be accounted for.

X-ray temperatures are in principle easier to understand: since the
ICM temperature is mainly determined by gravitational processes, it
should be more directly related to the total gravitating mass of the
cluster, i.e.~the $M$--$T$ relation should have less scatter and be
more tractable theoretically.  Although this is true to a first
approximation, the simplest expectation based on hydrostatic
equilibrium and isothermal gas have been shown to provide a poor
representation of the observed $M$--$T$ relation (e.g.~Finoguenov et
al.~2001; Allen, Schmidt \& Fabian 2001).  In the absence of improved
theoretical modeling one is forced to rely on an empirical
calibration of the $M$--$T$ relation, which raises many of the same
issues as for the $M$--$L$ relation discussed above.

In general, all methods of mass estimation have both a significant scatter
and potential systematic errors, leading to discrepant $\sigma_8$ results.

For example, Pierpaoli, Scott \& White (2001, hereafter PSW) explored several
improvements in determining $\sigma_8$ from the distribution of local
cluster temperatures, using a combination of X-ray data, cluster
simulations and theoretical modeling. For $\Omega_{\rm m}=0.3$, they
constrained the power spectrum normalization to lie around
$\sigma_8\simeq 1$, but stressed how existing uncertainties,
particularly in the Mass--Temperature relation, dominate the error bar.

Borgani et al.~(2001; hereafter B01, see also Rosati et al.~2002)
analyzed the luminosity distribution for the {\sl ROSAT\/} Deep
Cluster Survey out to $z\sim 1$.  They paid particular attention to
the impact of uncertainties and scatter in the $M$--$L$ conversion on
the final results.  They preferred a lower normalization,
$\sigma_8\simeq 0.7$ for $\Omega_{\rm m}=0.3$, with a roughly 15 per
cent uncertainty mainly driven by the systematics affecting the
$M$--$L$ relation.

In this paper we use the local XLF from the REFLEX survey
(B\"ohinger et al.~2002) and the distribution of temperatures from a
new compilation of nearby clusters

based on HIFLUGCS (Reiprich \& B\"ohringer~2002). Besides providing further
constraints on $\sigma_8$, we will discuss how these two different
approaches to the cluster mass function are affected by
systematics, and under which conditions they provide consistent
constraints on cosmological parameters. Our basic conclusion will be
that the two approaches in fact give consistent results, although at
present it is not realistic to expect a determination of $\sigma_8$ to
better than 10 per cent.  

More specifically, after giving the theoretical frame (Section
\ref{sec:theory}), we analyze the temperature-derived constraint in
Section~3.  We assess several possible different sources of error,
including: the $M$--$T$ normalization and its scaling; the effect of
the $L$--$T$ normalization error on the computation of the effective
volume; different temperature determinations; and different cuts of the
data.  For luminosity-derived constraints, in Section~4 we compare the
results obtained from the local REFLEX sample with the RDCS, therefore
testing possible evolutionary effects.  We also compare results
obtained by applying the direct $L$--$M$ relation with those obtained
via the the $L$--$T$--$M$ approach. In the end, we also compare the
luminosity- and temperature-derived constraints. We thoroughly
discuss how error treatments can influence the final result
throughout, particularly in Section~5, where we also compare our
results to other recent determinations of $\sigma_8$. Finally, we
present our conclusions in Section~6.

\section{Theory} \label{sec:theory}

The mass variance is constrained from the cluster abundance through
the mass function -- the (comoving) number density of objects of a
specified mass.
While there has been significant progress in the theory of the mass function
of late, there still remain serious systematic uncertainties in the
theoretical predictions, and it is therefore important to specify which
expression for the the mass function is being adopted.
Jenkins et al.~(2001) used a variety of $N$-body simulations for
different cosmological models to derive analytic expressions for the
mass function of dark matter halos. In particular, the expression in their
equation~(B3),
\be
{dn\over dM}\,=\,0.301\,{\bar\rho \over
M^2}\,{d\ln{\sigma^{-1}}\over d\ln{M}}\,
\exp\left(-|\ln{\sigma^{-1}}+0.64|^{3.82}\right),
\label{eq:jen}
\ee 
has been shown to reproduce the distribution of halos, if the mass is
interpreted as that within the radius, $r_{180{\rm m}}$, interior to which the
mean overdensity is $\rho=180 {\bar\rho_{\rm m}}$, for their `$\tau$CDM'
simulation.  Here $\bar\rho_{\rm m}$ is the `background' density or the mean
cosmic mass density (i.e.~$\Omega_{\rm m}$ times the critical density).
They also show that the mass function is approximately `universal' if the
mass is taken to be the sum of the particles found in their simulation with
a particular group finder.
The combination of these findings strongly suggests that equation~(\ref{eq:jen})
describes the mass function for a wide range of theories if the mass is
measured interior to $r_{180{\rm m}}$ (see also Hu \& Kravtsov 2002;
White 2002).

Jenkins et al.~(2001) have confirmed and extended earlier work which shows
that the standard Press--Schechter expression for the mass function
(Press \& Schechter 1974) significantly overestimates and underestimates the
number density of halos in the low- and high-mass end of their
distribution, respectively (see also Evrard et al.~2002).
They also found that the mass-function by Sheth \& Tormen (1999) somewhat
overestimates the number density of very massive halos, although this
difference is probably negligible for our current purposes.
However, in general, different expression for the mass functions induce
differences of 4--8 per cent in the resulting value of $\sigma_8$, which are
comparable to or larger than the statistical uncertainties.

In addition to this, Jenkins et al.~(2001) and White (2002) have shown
that the mass function is not precisely a universal function,
i.e.~it is not simply
a function of the linear density field smoothed on an appropriate scale.
While different cosmologies predict mass functions which are very similar,
in scaled units, there is a `scatter' of several tens of per cent in number
at fixed mass.
The non-universality of the mass function introduces an additional, and
non-negligible, systematic uncertainty in the conversion between a measured
object's abundance and the cosmological parameters.  As we push to more
precise estimates of increasing numbers of parameters these issues will need
to be confronted head on.  However, for the moment we shall ignore such
refinements as our focus will be on currently more important systematic errors.

The mass entering in equation~(\ref{eq:jen}) should be interpreted as the mass
contained within $r_{180{\rm m}}$, i.e.~inside a radius
encompassing a mean overdensity $\rho=180 \bar\rho_{\rm m}$.
However, scaling relations connecting mass to X-ray observable quantities
may provide the mass at different values of $\rho/ {\bar\rho}$
(see below).
In this case we follow White (2001) and rescale the masses assuming an NFW
profile for the DM halo (Navarro, Frenk \& White 1996) with a concentration
$c=5$ appropriate for a rich cluster.  A similar procedure by
Hu \& Kravtsov (2002) gives essentially identical results.

The quantity $\sigma(M,z)$ in equation~(\ref{eq:jen}) describes the
perturbation of the mass-scale $M$, which is the mass contained
within a top-hat sphere of comoving radius $R$, such that
$M=(4\pi/3)\bar{\rho}R^3$. In terms of the power spectrum, it is
\begin{equation}
  \si^2(R,z)=\int_0^\infty {dk\over k}\ \Delta^2(k,z)\, W^2(kR)\,,
\label{eq:sir}   
\end{equation}
where $\Delta^2 = k^3 P(k,z)/(2\pi^2)$,
$P(k)\equiv |\delta_k|^2 \propto k^n T^2(k)$
is the matter power spectrum ($n=1$ will be assumed in the
following for the primordial spectral index), $W(kR)$ is the window
function corresponding to the smoothing of the density field (the
Fourier transform of a top-hat in this case) and $T(k)$ is the
transfer function.  For the latter we use the fitting expression
provided by Eisenstein \& Hu (1999).

We notice here that some commonly used approximations for $\si_8$ or
for the transfer function may introduce systematic
sources of error.
For example, the  Viana \& Liddle fitting formula for $\sigma_8$
as a function of cosmology is imprecise at the level of 2--3 per cent
for $\Omega_{\rm m}$ values as low as 0.2.
Another common choice is the use of the Bardeen et al.~\shortcite{BBKS}
fitting formula with no baryons in conjunction with an expression for
the `shape parameter' $\Gamma$ which models the baryon dependence. 
This can introduce differences in the transfer function of 5--20 per cent
around $k \simeq 0.1\, h\,{\rm Mpc}^{-1}$.
We will stick with the Jenkins et al.~(2001) mass function and the
transfer function given by Eisenstein \& Hu (1999).

In carrying out these comparisons, we restrict ourselves to flat
cosmological models with an initial power spectrum index of $n=1$, and
pure cosmological constant making up the dark energy. 
We will also fix the baryon fraction to be $\Omega_{\rm b}h^2=0.02$,
and either allow $\Omega_{\rm m}$ to be free, or fix its value at 0.3.
Scaling relations for other cosmologies can be found in PSW.

Once one has adopted a cosmology and a mass function, then the power
spectrum can be normalized using some property of the clusters which
can be used to estimate mass.  For studies which focus on the X-ray
properties of clusters, then all methods effectively use either the
X-ray temperature function (XTF) or the X-ray luminosity function
(XLF).

\section{Constraints from the XTF} \label{sec:xtf}

In this section we present our $\si_8$ determination derived from the
temperature measurements.  We first discuss the sample used
(Section \ref{sec:local}), then
the possible choices for temperature modeling (Section \ref{sec:temp}), the 
$M$--$T$ scaling relation (Section~\ref{sec:MT}), our methodology 
(Section~\ref{sec:meth}), and finally the results (Section~\ref{sec:resT}).

\subsection{The local sample} \label{sec:local}

Since almost all recently used cluster samples derive from essentially the
same {\sl ROSAT\/} databases, it is intriguing that different selections and
treatment of the data apparently give such different results.  It is therefore
clearly important to define one's sample as carefully as possible.

The error bars on $\sigma_8$ are not dominated by Poisson fluctuations in
the small number of sources, but by systematic sources of scatter.  So
it is important to have a well understood cluster sample, rather than
the largest available one.  We specifically focus on a
{\it local sample}, by which we mean clusters with red shift $z \le 0.1$.
This allows us to avoid evolution issues. 
We also exclude the nearest clusters (at
$z<0.03$) to avoid possible biases introduced by survey incompleteness
and local super-clustering.

In PSW we defined a cluster catalogue adapted from the approach of
Markevitch \shortcite{Markevitch}, which was based on {\sl
ROSAT}-selected clusters with {\sl ASCA}-derived temperatures.  Our
temperatures there were taken from White \shortcite{DWhite}, who
fitted {\sl ASCA\/} data with a model which included the effects of
cooling flows.  We supplemented these cooling-flow corrected
temperatures with a few temperature estimates drawn from the
literature.  Since that work the available X-ray catalogues have
improved significantly.

The HIFLUGCS catalogue \cite{ReiBoh02} has now been published, and is
probably the most complete publicly available X-ray catalogue based on
the {\sl ROSAT\/} All Sky Survey (RASS, Tr\"umper et al.~1990).
Reiprich \& B{\"o}hringer (2002, hereafter RB02) discuss the selection
of the HIFLUGCS sample in detail.  Here we give the most relevant
features only.  The main selection criterion is a flux threshold at
$1.7\times10^{-14}{\rm W}\,{\rm m}^{-2}$ in the {\sl ROSAT\/}
$0.1$--$2.4\,$keV band.  The final flux limit is set at $f_{\rm
lim}=2.0\times10^{-14}{\rm W}\,{\rm m}^{-2}$, above which it appears
the sample is very close to complete over the $8.14\,$sr it covers.
This sample contains 63 clusters, with an additional 43 clusters in an
extended list.  Rather than use the {\sl ASCA\/} derived, cooling flow
corrected temperatures from White \shortcite{DWhite}, we have used the
temperatures quoted by Ikebe et al.~\shortcite{Ike01}.  They present
temperatures for the HIFLUGCS clusters determined from a
two-temperature fit.

In constructing our sample we exclude both the nearest
($z\,{<}\,0.03$) and most distant ($z\,{>}\,0.1$) clusters, to avoid
possible biases.  Within these limits we wish to construct an
effectively temperature limited sample, so we need to adopt a relation
between X-ray temperature and luminosity.  The luminosity-temperature
relation for HIFLUGCS is (Ikebe et al.~2002, equation~(4))
\begin{equation}
  L = (1.38\times10^{35})\, T^{2.5}\, h^{-2}\, {\rm W},
\end{equation}
with $T$ measured in keV and $L$ in the 0.1--2.4 keV energy band.

We include a cluster if it would have passed the flux cut had its
luminosity been given by $L(T)$ and if it is hotter than $3\,$keV.
The errors on the measured temperature are included by weighting each cluster
based on the probability it is in the sample assuming a Gaussian distribution
of temperatures described by the observed central temperature and error.

While we use the mean $L$--$T$ relation to compute the weight of each cluster,
when we later compute the effective distance to which a cluster of temperature
$T$ could have been seen, we include the effect of scatter in $L$--$T$
(which slightly reduces the distance).

In the following we will use the primary HIFLUGCS sample  (63 clusters)
as the reference one, but we will also compare the results derived using
the PSW catalog (72 clusters) and an extended version of the HIFLUGCS
sample which contains 96 clusters from a combination of the extended HIFLUGCS
and PSW.
Note that after applying the selection criteria presented above, the actual
number of clusters used to determine $\sigma_8$ is significantly lower (see 
Section~\ref{sec:resT}).

\subsection{Modeling cluster temperatures} \label{sec:temp}

Authors differ in what they mean by cluster `temperature' and there are a
number of subjective choices which can be possible sources of
discrepancy, including: different models for fitting $T$; varying
fields of view and spatial resolutions; spectral band differences;
and methods for dealing with sub-structure, modeling mergers, etc.
The basic point is that the quoted errors do not include all of the systematic
effects and care must be taken to avoid comparing apples with oranges.

A particular difficulty is deciding how to deal with radial structure,
particularly in the cluster center.
Clearly some procedure needs to be adopted, since a single isothermal
model does not adequately fit both the central and outer regions of many
clusters.  So long as the same procedure is adopted for the method used
to fix the $M$--$T$ normalization as is used for the data,
it probably makes little difference precisely what is done.
However this is rarely the case in practice.
In PSW, for example, we used temperatures from
White \shortcite{DWhite} which came from a specific cooling flow
model which had been fitted to {\sl ASCA\/} data, while
Markevitch \shortcite{Markevitch} simply excised the centres of clusters.
The justification for this is that most simulations possess neither the
resolution nor the physics to adequately model cooling flows, and hence
they should not be allowed to affect the data either.

Recently Ikebe et al.~\shortcite{Ike01} performed a different fitting
for the temperatures of the clusters.  They use a 2-temperature
picture, in which the outer regions are modeled by a single
temperature isothermal plasma, while in the central regions there is a
second cooler component forming a multi-phase plasma with the first.
For some clusters, with no statistical measurement of a cooler
component, they fixed its temperature at $T_{\rm hot}/2$.  They argue
that overall this method provides a reasonable description for both
cooling-flow and non-cooling-flow clusters.  We will therefore use the
temperature of the hot isothermal gas as the mass proxy for each
cluster in our sample.  It is worth pointing out that the
temperatures derived using this method are typically lower than those
obtained by some previous authors (see fig.~1 in Ikebe et al.~2002);
in particular, the discrepancy is bigger for hotter clusters.  For
clusters hotter than $6\,$keV, the White \shortcite{DWhite}
temperatures are approximately 25 per cent hotter than those of Ikebe
et al.~\shortcite{Ike01}.  It now seems likely that the White~(2000)
temperatures were overestimated in several cases.

\begin{figure}
\begin{center}
\leavevmode
\epsfxsize=8cm \epsfbox{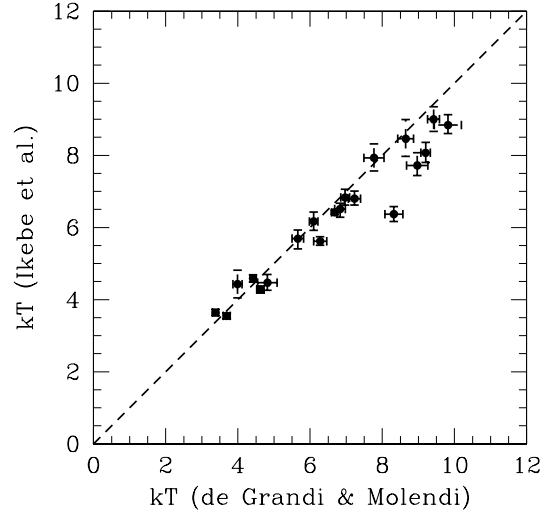}
\end{center}
\caption{Ikebe et al.~(2002) isothermal temperature fits from {\sl
ASCA\/} data compared with de Grandi \& Molendi~(2001) temperatures
derived from {\sl Beppo-SAX\/} data for those clusters in common.
This is indicative of the current size of systematic errors in
temperature estimates.}
\label{fig:Tcomp}
\end{figure}

However, to further illustrate this issue we plot in
Fig.~\ref{fig:Tcomp} a comparison of Ikebe et al.~\shortcite{Ike01}
temperatures with those which are in common with a study of clusters
using {\sl Beppo-SAX\/} data by de Grandi \& Molendi
\shortcite{deGMol}.  Firstly, it is clear that these entirely
independent determinations of temperature are in rather good
agreement.  However, in detail there are differences for individual
clusters, and a general trend for the Ikebe et al.~\shortcite{Ike01}
temperatures to be cooler for the highest $T$ objects.  This
comparison should serve as a useful guide to the size of possible
systematic effects in the temperature determinations.  We will later
consider the effect on our $\sigma_8$ estimate of boosting the
temperatures of the highest temperature clusters.

In the following we will adopt temperatures from Ikebe et al.~(2002)
whenever available, that is for the main HIFLUGCS sample, and the HIFLUGCS
extended sample, while we adopt temperatures from PSW
for the PSW sample and for the clusters in that supplementary sample.

\subsection{The Mass--Temperature relation} \label{sec:MT}

The $M$--$T$ relation is generally assumed to be:
\begin{eqnarray}
  \left( {M(T,z)\over 10^{15}\,h^{-1}\,{\rm M}_\odot }\right)
&=& \left( {T\over{T_\ast}} \right)^{3/2}
    \left(\Dc E^2 \right)^{-1/2} \times \nonumber \\
& & \left( 1 -  2  {\oml(z)\over\Dc  }\right)^{-3/2}\ ,
\label{eqn:m-t}
\end{eqnarray}
where $T$ is in keV\footnote{In PSW we called this normalization
$\beta$, following some
of the simulation papers.  However, this caused some confusion because a
different subset of authors (motivated by the isothermal $\beta$ model) have
defined a $\beta$ which is inversely proportional.  In addition $\beta$ is
often used to denote the outer slope of the emission profile of clusters.
Defining $T_\ast$ in this way is much less ambiguous.},
$\Dc$ is the mean overdensity inside the virial radius
in units of the critical density and
$E^2=\omm (1+z)^3 + \oml + \Omega_{\rm k} (1+z)^2$.
Note that $\Dc$ depends on cosmology, with
fitting formulae given in PSW, and
is a redshift dependent variable, which should be evaluated
using the appropriate $\oml(z)$ and $\omm(z)$.

While equation~(\ref{eqn:m-t}) comes from rather simplistic arguments
(essentially
dimensional analysis and an assumption that clusters are self-similar) both
observations and simulations suggest it is a good approximation for systems
hotter than about $3\,$keV.  For example Finoguenov et al.~(2001) find
\begin{equation}
  M_{500} = (4.22^{+0.85}_{-0.66})\times 10^{13} h_{50}^{-1}{\rm M}_{\odot}
 \times kT^{{1.48}^{+0.10}_{-0.12}},
\label{eq:finoMT}
\end{equation}
for $kT>3\,$keV clusters, where $M_{500}$ is the mass within an
overdensity of 500 with respect to critical.
Other recent determinations are also consistent with a slope of 1.5
(e.g.~Allen et a.~2001; Xu et al.~2001)
provided that only the higher temperature clusters
are considered.

Unfortunately the level of agreement on the overall normalization,
$T_\ast$, is not so good (see e.g.~Table 1 of Muanwong et al.~2002
or fig.~2 of Huterer \& White 2002).
In fact the uncertainty in $T_\ast$ has for some time been the dominant
uncertainty in determining $\sigma_8$.
In PSW $T_\ast$ was assumed to be 1.3, in B01 it was assumed to be 1.2.
These values are representative of the older simulation results.
The observations have almost always suggested a lower normalization for the
$M$--$T$ relation than implied by the theoretical model (i.e.~a higher
value of $T_\ast$), which therefore leads to a lower $\sigma_8$ value.
A rough scaling argument suggests (Huterer \& White 2002, see also Fig.~2)
\begin{equation}
  \Omega_{\rm m}^{0.6} \sigma_8 \propto \left(T_\ast\right)^{-0.8},
\end{equation}
with a slightly weaker dependence if only the most massive clusters are used
in the analysis (Evrard et al.~2002).
Including additional physics in the simulations and trying to estimate the
mass in the same way as the observers helps to reconcile the discrepant
normalizations (e.g.~Muanwong et al.~2002) and suggests that $T_\ast$ may
be closer to 1.6 than the values assumed earlier. 
Some direct observational determinations (which estimate the mass in
different ways) suggest an even higher value
(Finoguenov et al. 2001, Xu et al. 2001).

We would like to stress that there are two kinds of uncertainty in the
$M$--$T$ relation: an intrinsic scatter in the temperatures of clusters
of a given mass which arises due to differences in their formation history;
and an overall uncertainty in the normalization of the $M$--$T$ relation.
The first sort of uncertainty is entirely statistical in character.
It needs to be considered, because it both moves the central value for
$\sigma_8$ and contributes to the error bar.  Taking this scatter into account
is straightforward if its distribution is known.  Most studies have agreed
that the scatter is around 10--15 per cent in $T$,
and is essentially a reflection of the different merging histories.
The second sort of uncertainty is entirely systematic, however, and has to
be treated in a different way, for example by marginalising the likelihood
of $\sigma_8$ over some prior distribution of values of $T_\ast$.
This distinction is not typically made in determinations of $\sigma_8$ from
X-ray data.

It is very easy for errors in the determination of $T_\ast$ to creep in.
Recall that we are ultimately interested in determining the
mass within a sphere of radius $r_{180{\rm m}}$
centered on the cluster in order to compare to theoretical mass functions.
For a typical rich cluster $r_{180{\rm m}}\simeq 2-3\, h^{-1}$Mpc,
well beyond the region where observations probe directly, 
with the possible exception of lensing studies. 

In
order to determine this mass we may estimate $M_{\Delta}$ at one of
various different density contrasts $\Delta$ using a variety of
methods each known to have biases.  Assuming an NFW profile with a
constant concentration ($c=5$ is assumed for cluster scale halos) and
a flat cosmology with $\omm=0.3$, then some commonly used masses are
related by
 $ M_{\rm vir} \equiv M_{\Dc} = 1.22\, M_{200}, 
  {\rm and} \ M_{500}  = 0.72\, M_{200}$, 
while $\Dc \simeq 100$.
Hence we can derive $M_{\rm vir}=1.69\, M_{500}$, so the $M$--$T$ relation of
equation~(\ref{eq:finoMT}) implies a normalization of $T_\ast\simeq 1.9$.
Significant corrections are being applied in these conversions, involving large
extrapolations from given profiles, indicating that there is plenty
of room for error.

We caution that there is also a significant scatter in cluster concentrations,
and that the NFW profile is only `universal' in that it fits statistically
to isolated clusters, rather than holding in detail for each cluster.
We believe that these issues do not have a large effect on determining the
cluster normalization \cite{MassFunction}, however this could be studied in
more detail.  Careful direct comparison between simulations and
observations is probably ultimately the best approach when the next generation
of simulations becomes  available.

In summary, while both observations and simulations agree quite well
on the scaling law of equation~(\ref{eqn:m-t}), there is no obvious
agreement on the value of the normalization $T_\ast$.  The
observations may favour a higher $T_\ast$, perhaps 1.8--1.9, while the
simulations at present appear to give values around 1.5--1.6. At
present it is by no means clear which should be the preferred value,
or indeed if they are genuinely discrepant.  In what follows, we will
investigate the implications on the determination of $\sigma_8$ of
adopting different $M$--$T$ assumptions and error treatments.

\subsection{Methodology} \label{sec:meth}

The general method is similar to the one described in PSW, with some
technical improvements.
We developed two independent codes for the likelihood evaluation, with
slightly different approaches for the treatment of the errors, and we
cross-checked the results.  This allows us to to identify areas of potential
disagreement in the $\si_8$ estimate.

Perhaps the largest modification is that our former Monte Carlo method for
the error treatment has been completely replaced by `weight functions' to
increase speed.
For each set of cosmological parameters, including the normalization and slope
of the mass temperature relation, we construct a grid of $n(T)dT$, taking into
account the asymmetric nature of the quoted observational errors on the
temperatures.  At this stage a volume cut based on the $L$--$T$ relation is
applied.  A conversion from temperature to mass, and between mass definitions,
is used to compare the observed $n(T)$ with theory using the Poisson averaged
likelihood
\begin{equation}
  \left\langle \log {\cal L} \right\rangle = \sum_i d_i \log \mu_i - \mu_i,
\label{eqn:loglikelihood}
\end{equation}
where $d_i$ is the fraction of the cluster in bin $i$ and $\mu_i$ is
the mean value computed from the mass function and effective volume.

In principle the uncertainty in the mass temperature relation could
be treated in two different ways: either with a convolution of the mass
function or by marginalising the value of the likelihood.
The former approach is appropriate for treating intrinsic scatter, while
the latter is more appropriate for an uncertain parameter such as the
overall normalization or slope.
In PSW we showed that the uncertainty in relating the mass to the observable
dominates the final error bar, so it is important to treat this issue as
carefully as possible.  Both approaches have been adopted in the literature,
accordingly, we investigated the influence of the two approaches on the
likelihood independently.
We also checked that the intrinsic scatter could be treated equivalently
through either convolution or by an extended Monte Carlo procedure.

\subsection{Results} \label{sec:resT}

We present here our results, quoted with respect to a reference model
which assumes the HIFLUGCS extended catalogue with temperature from
Ikebe et al.~(2002). Unless otherwise stated, we will focus on
$T_\ast=1.75$ and $\Omega_{\rm m}h = 0.2$. 
Our fiducial choice for
scatter in the $M$--$T$ relation is 10 per cent in $T$, with a
Gaussian distribution.

\begin{figure} 
\centerline{ \hbox{
\psfig{file=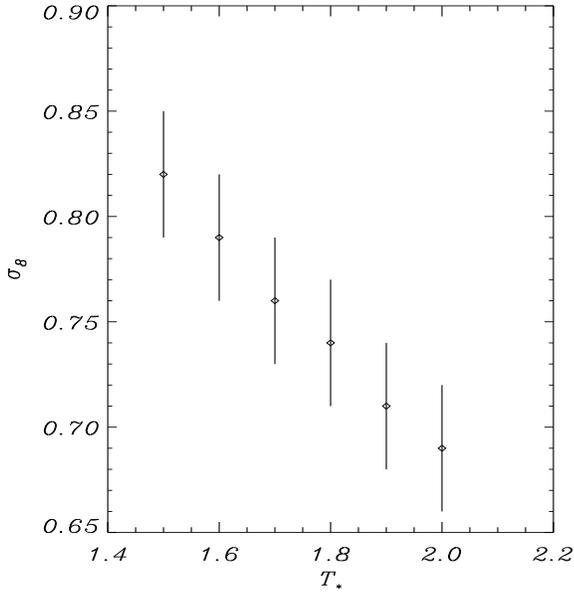,height=8.5cm,width=8.5cm,angle=0} } }
\caption{ The dependence of $\si_8$ on the value of the $M$--$T$
normalization $T_\ast$ assumed.  Here we fix $\Omega_{\rm m}=0.3$ and
$h=0.7$.  Error bars are $1 \sigma$ assuming an intrinsic scatter
of 10 per cent in $T_\ast$.  The choice of $T_\ast$ is probably the most
important difference between some discrepant values of $\sigma_8$ in the
literature.}  
\label{fi:s8vsbe} 
\end{figure}

We show in Fig.~\ref{fi:s8vsbe} the variation of $\sigma_8$ with $T_\ast$ 
if we hold $\Omega_{\rm m}=0.3$.
We find that the uncertainty due to the choice of the  fiducial $T_\ast$ 
value dominates with respect to the assumed intrinsic scatter.
In order for this situation to be improved in future studies, 
the discrepancy between the observational and simulated $T_\ast$ values
will need to come into better agreement.

\begin{figure} 
\centerline{ \hbox{
\psfig{file=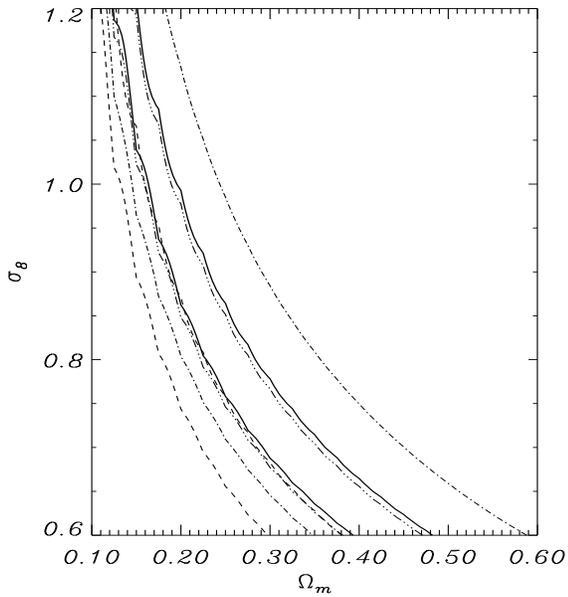,height=8.5cm,width=8.5cm,angle=0} } }
\caption{ 90 per cent confidence regions in the $(\sigma_8,\Omega_{\rm
M})$ plane from the analysis of cluster temperatures.  The solid
contour correspond to a fixed $T_\ast=1.75$ (no $T$ scatter), the
dot-dashed is obtained by convolving the likelihood function with a 10
per cent error in $T$, while the three-dots-dashed and dashed are
obtained by convolving the mass function with a Gaussian window with a
10 per cent and 16 per cent width in $T$.}
\label{fi:licoma} 
\end{figure}

In Fig.~\ref{fi:licoma} we show the effect of treating the uncertainty in
$T_\ast$ in two different ways: via convolution of the likelihood function
or of the mass function.
It is apparent that after marginalisation of the likelihood is performed,
the overall effect is to broaden the error bars with respect to the case
with no errors.  Convolution of the mass function,
 instead, mainly shifts $\si_8$ toward lower values.
The two approaches are therefore not equivalent, and should be used 
appropriately (i.e.~convolution or Monte Carlo for the intrinsic scatter
and marginalisation for the systematic normalization uncertainty).

These results have been obtained with Gaussian scatter in $T$.
Some authors (e.g. B01) have considered the scatter instead to be
Gaussian in $\log(T)$, as observed scaling relations sometimes
quote.  While there is no particular reason for the latter choice,
apart from the convenience of fitting a straight line to the data, we
verified that the two different choices give similar error bars, at
least at the $1\sigma$ level, and for the 10 per cent scatter assumed
here.  However, we caution that in practice the distribution may not
be Gaussian, and the tails may have a disproportionate effect.  We
also note in Fig.~\ref{fi:licoma} that for relatively small intrinsic
error (10 per cent in $T$) the effect of the scatter is sub-dominant
with respect to the other sources of error.  However, we see that a
16 per cent scatter is sufficient to shift the $\sigma_8$ value
by a significant amount.

\begin{figure} 
\centerline{ \hbox{
\psfig{file=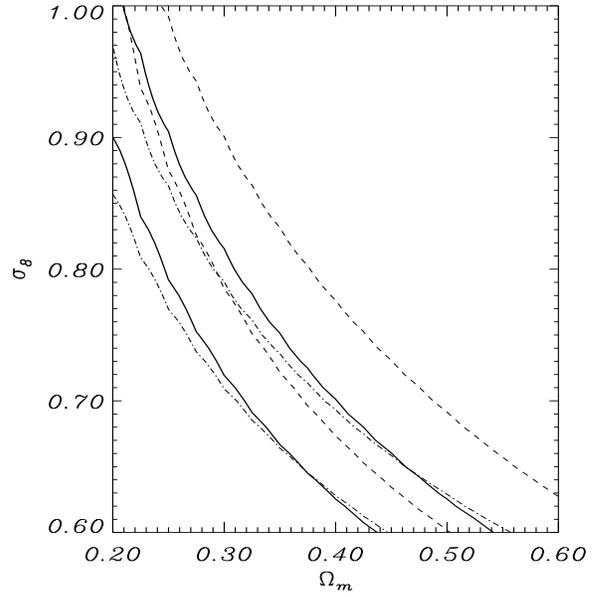,height=8.5cm,width=8.5cm,angle=0} } }
\caption{The 90 per cent confidence contours obtained with different
mass functions for a fixed $T_\ast =1.75 $. The solid line corresponds
to the Jenkins et al.~(2001) mass function in equation~(\ref{eq:jen}),
the dashed line to that of Sheth \& Tormen (1999), and the dot-dashed
to the Press \& Schechter (1974) formula. }
\label{fi:limf} 
\end{figure}

We show in Fig.~\ref{fi:limf} the dependence on the assumed mass
function.  We note here that the results derived with the Jenkins at
al.~(2001) mass function are more similar to the ones obtained using
Press--Schechter~(1974) than to those assuming the Sheth \& Tormen
(1999) form.  This is due to the fact that the Jenkins et al.~(2001)
function is actually closer to Press--Schechter than to Sheth--Tormen
in the range $0.3 < \ln{\sigma^{-1}} < 0.7$ which is mostly relevant
here.

\begin{figure} 
\centerline{ \hbox{
\psfig{file=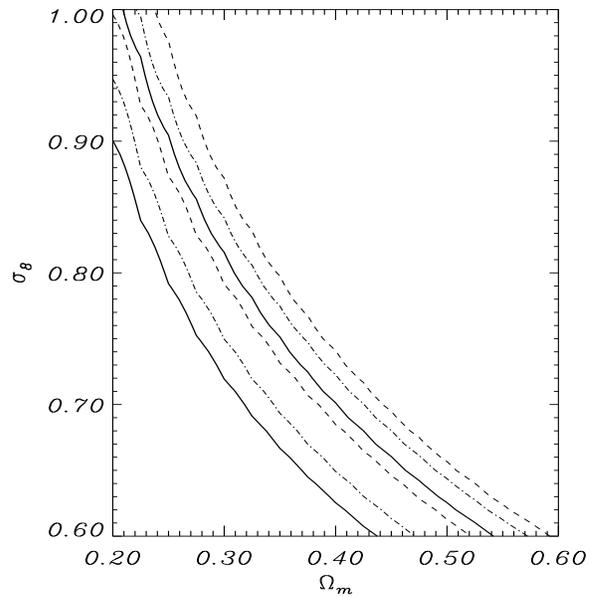,height=8.5cm,width=8.5cm,angle=0} } }
\caption{The 90 per cent confidence contours obtained using different
input catalogues for a fixed $T_\ast =1.75 $. The solid line is
obtained using the HIFLUGCS sample, the dot-dashed with the extended
version (the most appropriate sample to use in our particular
methodology), and the dashed with the old PSW catalogue. In the latter
case, the $L$--$T$ relation used is the one from PSW.}
\label{fi:cata} 
\end{figure}

The dependence on the specific sample used is shown in
Fig.~\ref{fi:cata}.  We note that although the cluster sample contains
97 clusters in its extended version, due to our selection criteria the
actual number of clusters used is much lower: the code selects on
average 28.5 clusters out of the HIFLUGCS sample, 38 out of the
extended one and 41.5 out of the old PSW one.  About 10 per cent
discrepancy between the HIFLUGCS sample and the PSW one is due to the
inclusion of the clusters that are nominally below the flux limit (and
scatter up from the extended sample).  These clusters therefore give a
non-negligible contribution to the final $\si_8$ value through our
likelihood technique.  The rest of the discrepancy may be explained
with the different modeling of the temperatures between Ikebe et
al.~(2002) and White~(2000).

For the dependence of $\si_8$ on cluster temperature, we verified that
cutting out clusters below $4\,$keV does not result in a significant
difference.  This means that our $3\,$keV low temperature cut is high
enough to include clusters that actually follow the same $M$--$T$
relation.  We also boosted the temperatures of these clusters above
$6.5\,$keV by 10 per cent, finding no noticeable change in the 90 per
cent contour levels.  This implies that the level of discrepancy in
temperature estimation between, for example, Ikebe et al.~(2002) and
de Grandi et al.~(2002) is probably not a major source of uncertainty.

We also tested the error in the slope of the $M$--$T$ relation,
finding that it is not very important.  Increasing the exponent from
1.5 to 1.6 lowers $\sigma_8$ by 0.05, while decreasing it to 1.4
increases $\sigma_8$ by a similar amount.  Since we define our pivot
point at $10^{15}\,h^{-1}{\rm M}_\odot$, above where most of the
sample lies, the sense of the shift is as expected.

Finally, introducing a 35 per cent scatter in the $L$--$T$ relation
when computing volume implies a reduction of about 8 per cent in the
range $4<T<8$ keV.  However, this has a negligible effect on the final
$\sigma_8$ result.

After investigating all of these effects, it is clear that one has to
be very careful in order to obtain a robust estimate for $\sigma_8$.
But even when that has been achieved, the overall error is dominated
by the systematic uncertainty in the normalization of the $M$--$T$
relation.

To obtain an overall `best efforts' estimate for $\sigma_8$
we can marginalise over a uniform distribution for $T_\ast$,
stretching between typical simulation values (1.5) to typical
observational values (1.9).  For the standard $\Lambda$CDM model (with
$\Omega_{\rm m}=0.3$, $h=0.7$, $\Omega_{\rm b} h^2=0.02$ and $n=1$) we
obtain
\begin{equation}
  \sigma_8=0.77^{+0.05}_{-0.04}
\end{equation}
(c.f.~Fig.~\ref{fi:s8vsbe}),
where the error bars describe the 68 per cent confidence region
(corresponding approximately to $\pm1\sigma$).  Here we use our best
attempt to construct an effectively temperature-selected sample, which
consists of the HIFLUGCS extended sample (with suitable cuts), plus
some additional clusters from PSW.  Scalings similar to those given in
PSW could be used to extend this to different cosmologies.

\section{Constraints from the local XLF} \label{sec:xlf}

An alternative approach is to dispense with the X-ray temperatures and
to use the X-ray luminosity function (XLF). This method suffers from
the potential limitation that the relation between X-ray luminosity
and mass is less well understood and is more affected by uncertainties
in the thermodynamic status of the ICM than the $M$--$T$ relation.
However, the XLF has the significant advantage that it is now
precisely calibrated for extended sets of both nearby and distant
($z\sim 1$) cluster samples.  
Quite remarkably, all the determinations of the local XLF based on the
{\sl ROSAT\/} All Sky Survey (RASS) and on deep pointings agree pretty
well with each other(see Rosati, Borgani \& Norman 2002, for a
review), thus providing a stable reference point to establish the
evolution of the X-ray cluster population.

Therefore, it is important to understand the extent to which XTF and
XLF approaches give consistent results on cosmological constraints.
This is only possible through a careful consideration of the different
steps involved.  The most recent determination of the local XLF is
based on the REFLEX survey (B\"ohringer et al.~2002, B02
hereafter). This survey includes about 450 clusters out to $z\mincir
0.3$, and is complete down to the flux limit of $3\times 10^{-15}{\rm
W}\,{\rm m}^{-2}$ in the 0.1--$2.4\,$keV energy band. B02 provided a
binned representation of the cluster XLF in the luminosity range
$L\simeq 4\times 10^{34}$ to $3\times 10^{39} {\rm W}$ (with $h=0.5$)
for both an Einstein-de Sitter cosmology and for a flat low-density
model with $\Omega_{\rm m}=0.3$. The effect of changing cosmology is
almost negligible at the low redshifts probed by the REFLEX
survey. For definiteness, we will use in the following the XLF
determination for the low-density model.

Schuecker et al.~(2002) combined clustering properties and redshift
distribution of REFLEX clusters, through a maximum-likelihood
approach, to obtain constraints on $\sigma_8$ and $\Omega_{\rm m}$.
They applied the $L$--$M$ relation calibrated by RB02 and obtained
$\sigma_8\simeq 0.75$ for $\Omega_{\rm m}=0.3$ with formal statistical
uncertainties of about 5 per cent. Consistent results have been also
found by Allen et al.~(2002), who used the XLF from the REFLEX survey
and the BCS (Ebeling et al.~2000), combined with an $L$--$M$ relation
calibrated from {\sl Chandra\/}/{\sl ROSAT\/} data and weak lensing
observations. Rather than focusing on a further determination of
cosmological constraints from the REFLEX XLF, our analysis is more
aimed at understanding how such constraints are affected by the
details of the analysis method, and how they compare with those
derived from the XTF analysis, presented in the previous section.

Since we use the binned representation of the XLF given by B02, we
determine best-fitting values for cosmological parameters and
confidence levels by computing the $\chi^2$ between the measured and
predicted REFLEX XLF.  The model XLF is computed from the mass
function of equation~(\ref{eq:jen}) by converting masses into observed
X-ray luminosities (see below).  After finding the values of the
$(\sigma_8,\Omega_{\rm m})$ parameters which minimize the $\chi^2$
value, confidence regions are estimated from standard 2-dimensional
$\Delta \chi^2$ variations.
We assume the REFLEX XLF to be determined at the effective redshift
$z_{\rm eff}=0.1$, which is close to the median redshift of the
survey.

\subsection{Converting masses into X-ray luminosities}

In order to convert the observed $L$ into mass to use in the
mass function of equation~(\ref{eq:jen}), we follow two different
procedures.

\subsubsection{Method (a): using $L$--$T$ and $M$--$T$}
\label{sec:LTTM}

The first method combines the observed $L$--$T$ relation and the $M$--$T$
relation calibrated from hydrodynamical cluster simulations. This procedure
has been applied by several authors in the analysis of flux-limited
cluster surveys (e.g.~Kitayama \& Suto 1998, Sadat, Blanchard \&
Oukbir 1998, Reichart et al.~1999, B01).
As for the $M$--$T$ relation, we use the expression of
equation~(\ref{eqn:m-t}). 

As for the $L$--$T$ relation, different analyses consistently show
that is well represented at cluster scales ($T\magcir 2\,$keV) by a
power law, $L\propto T^\alpha$, with $\alpha\simeq 3$, and with a
scatter which is significantly reduced once the effect of cooling
flows has been corrected for (e.g.~Markevitch 1998, Allen \& Fabian
1998) or by considering only systems without significant cooling-flow
signatures (e.g. Arnaud \& Evrard 1999).
Following the same notation for the relation between bolometric luminosity
and emission-weighted temperature as in B01, the expression we use is
\be 
L_{\rm bol}=L_6(T/6\,{\rm keV})^\alpha\,10^{37}h^{-2}\,{\rm W}\,,
\label{eq:lt} 
\ee 
with $L_6=3$ and $\alpha=3$.  Bolometric luminosity is then converted
into the 0.1--$2.4\,$keV band by using a {\sc mekal} spectral
synthesis code, assuming one-third solar metallicity for the
intra-cluster medium. Since this conversion has a larger effect for
hotter systems, the $L$--$T$ relation tends to flatten when computed
in the 0.1--2.4 keV band. This is the reason for the different slopes
in equations~(3) and (\ref{eq:lt}).
We assume that the combination of $M$--$T$ and $L$--$T$ variation
results in an overall 45 per cent scatter, which is assumed to be
Gaussian-distributed in the log. 

\subsubsection{Method (b): using $M$--$L$ directly}
\label{sec:ML}

The second method uses an observationally determined $M$--$L$ relation
(RB02, Ettori et al.~2002). RB02 estimated the $M$--$L$ relation for a
sample of 103 clusters with $L\magcir 10^{35}h_{50}^{-2}{\rm W}$,
temperature measured from ASCA and gas density profiles from {\sl
ROSAT\/} PSPC pointed observations. Ettori et al.~(2002) used {\sl
Beppo-SAX\/} data for a set of 22 clusters with $T>3$ keV, with
measured profiles of gas temperature and density. Based on such data,
they found that, unlike for the $L$--$T$ relation, no segregation
between cooling-flow and non cooling-flow clusters exists in the
$M$--$L$ plane. As a consequence, these authors claim that directly
measuring the $M$--$L$ relation is more stable than combining the
$L$--$T$ and $M$--$T$ relations.

In the following, we will use the $M$--$L$ relation from RB02, which
includes lower-luminosity systems, thus better covering the $L$ range
probed by the REFLEX XLF. We consider the relation based on mass
estimated at overdensity $\Delta=500$ and luminosity in the
$0.1$--$2.4$ band. As in Section~3.3, $M_{500}$ will be related to
$M_{\Delta_c}$ assuming an NFW profile with $c=5$ for the
concentration parameter.  RB02 claims that no extrapolation is
required for 86 per cent of the clusters in their sample to obtain
mass at such an overdensity (although obviously the assumption of a
$c=5$ NFW profile is still required to connect $M_{500}$ to
$M_{180{\rm m}}$).  RB02 adopted a fitting expression of the form
$L=10^A\left(M_{500}/h_{50}^{-1}{\rm M}_\odot\right)^\alpha
h_{50}^{-2} 10^{33}{\rm W}$ and provided fitting parameters for both
the direct ($A=-19.708$, $\alpha=1.652$) and inverse relations
($A=-17.545$, $\alpha=1.504$; see Table 10 in RB02). In the following
we will show results based on these different choices for $A$ and
$\alpha$, so as to judge the stability of constraints on cosmological
parameters against uncertainties in the fitting procedure.
%

Finally, RB02 also estimated the overall scatter in mass from this
relation to be $\simeq 48$ per cent.  After subtracting in quadrature
the contribution from the mean mass measurement uncertainty, $\simeq
28$ per cent, the resulting intrinsic scatter in the $M$--$L$ relation
turns out to be 39 per cent. 
This value overestimates the true scatter if errors in the mass
measurements have been underestimated, for instance due to the
assumptions of isothermal gas and $\beta$--model fitting for the gas
density profile, made by RB02 in their analysis.  Also, if significant
temperature gradients are present (e.g.~Markevitch et al.~1998,
Finoguenov et al.~2001, De Grandi \& Molendi 2002, Pratt \& Arnaud
2002), then the isothermal assumption would lead to an overestimate of
the total gravitating mass and, therefore, to an overestimate of
$\sigma_8$ at fixed $\Omega_{\rm m}$.  Whether violations of these
assumptions would lead to a bias or to an increase in the scatter or
both is unclear.

\begin{figure} 
\centerline{ \hbox{
\psfig{file=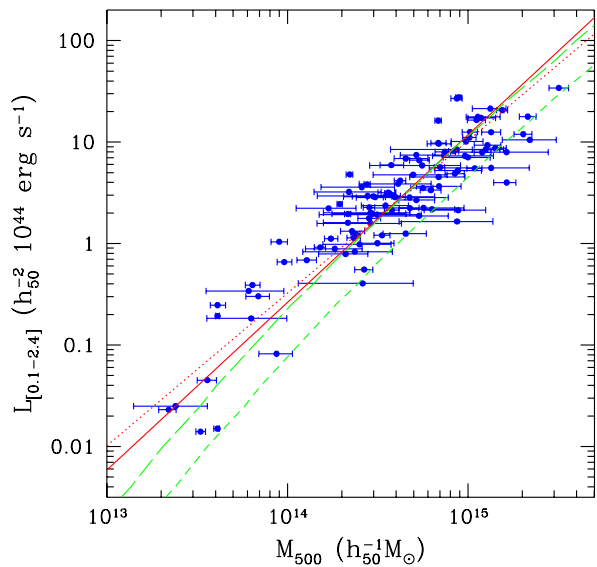,height=8.5cm,width=8.5cm,angle=0} } }
\caption{The relation between $M_{500}$ and
luminosity in the 0.1--$2.4\,$keV energy band. Data points are 
from Reiprich \& B\"ohringer (2002, RB02). Solid and dotted
lines represent the direct and inverse best-fitting relations from
RB02. Long-dashed and short-dashed lines have been obtained from the
combined $L$--$T$ and $T$--$M$ method to convert mass into $L$, after
assuming $T_\ast=1.75$ and $T_\ast=1.2$, respectively.  We assume
$\Omega_{\rm m}=0.3$ and use equation~(4) for the $M$--$T$ conversion.
Bolometric luminosities have been converted into $L$ in the
0.1--$2.4\,$keV band by using a {\sc mekal} model with $Z=0.3\,Z_\odot$
for the ICM metallicity.}
\label{fi:ml} 
\end{figure}

Following the same line of reasoning as in Section 3.4 for the
$M$--$T$ relation, any uncertainty in the $M$--$L$ relation can be
interpreted as due either to a genuine intrinsic scatter or to an
overall uncertainty in the normalization of the relation. In the first
case, one has to convolve the model mass function with a scatter,
while in the second case the correct procedure is to marginalise over
the amplitude of the $M$--$L$ relation, after assuming a range of
variation and a distribution for this parameter.
In the following we will assume as our fiducial analysis that based on
the RB02 $M$--$L$ relation, with 20 per cent intrinsic scatter in
mass.  We will also verify how the results change by marginalising
over the amplitude of the $M$--$L$ relation and by changing the
intrinsic scatter between the two extreme assumptions of negligible
scatter and 39 per cent scatter.  We show in Fig.~\ref{fi:ml} the
observational data points from RB02 along with the different $M$--$L$
relations that we use in our analysis.  Method {\em (a)\/} with
$T_\ast=1.2$ (used by B01) overestimates the mass at a fixed
luminosity, while taking $T_\ast=1.75$ is rather close to the fitting
relations used by RB02 for $L\magcir 10^{36}{\rm W}$.

\subsection{Results}
>From their analysis of the {\sl ROSAT\/} Deep Cluster Survey (RDCS),
B01 derived constraints on the $(\Omega_{\rm m},\sigma_8)$ plane by
following the evolution of the cluster population out to $z\magcir
0.8$ (see also Rosati et al.~2002). This analysis was based on the
combined $L$--$T$ and $T$--$M$ approach of method $(a)$ with
$T_\ast=1.2$, no significant evolution of the $L$--$T$ relation
(e.g.~B01, Holden et al.~2002, Novicki, Sornig \& Henry 2002; cf. also
Vikhlinin et al.~2002), an overall 45 per cent scatter in the $M$--$L$
relation, and assuming the mass function by Sheth \& Tormen (1999).

For the sake of direct comparison, we repeat here the same analysis on
the REFLEX XLF. We emphasize that a major difference between the analysis
of RDCS and of REFLEX lies in the widely different volumes covered by the
two surveys at different redshifts -- unlike REFLEX, the RDCS has a modest
volume coverage at low redshift, while being able to trace the cluster
population out to $z\sim1$. Therefore, the analysis of the RDCS
provides a {\em dynamical\/} constraint on $\Omega_{\rm m}$, by probing
the growth rate of density perturbations, while REFLEX provides a
{\em geometrical\/} constraint on $\Omega_{\rm m}$ through the shape of
the power spectrum (assuming CDM and fixing $h$ and $\Omega_{\rm b}$ from
independent observations).

\begin{figure} 
\centerline{ \hbox{
\psfig{file=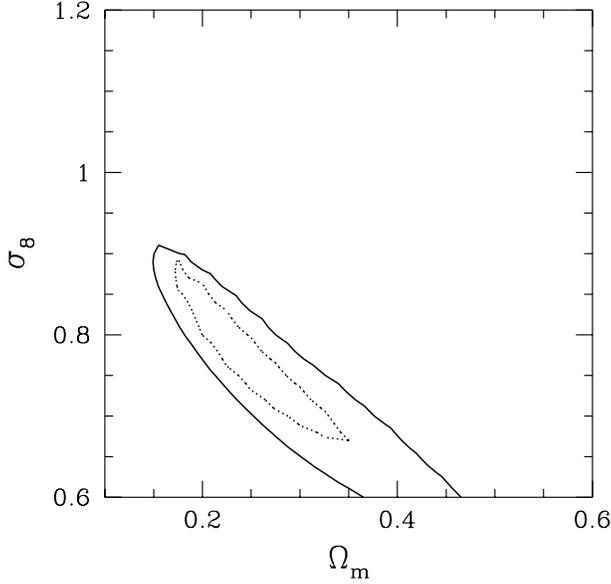,height=8.5cm,width=8.5cm,angle=0} } }

\caption{Comparison between the constraints derived from the REFLEX
  and the RDCS XLF. The solid contour indicates the 90 per cent
  confidence region from the REFLEX XLF, by using the method $(a)$ for
  the mass--luminosity conversion (Section~4.1.1), with $L$--$T$
  relation given by equation~(\ref{eq:lt}), $M$--$T$ relation by
  equation~(4) with $T_\ast=1.2$. The dotted contour corresponds to
  the analysis of the evolution of the RDCS XLF out to $z\simeq 0.8$
  (B01, Rosati et al.~2002; see text), for the same choice of $L$--$T$
  and $M$--$T$ relations and assuming a non-evolving $L$--$T$ (for
  $\Omega_{\rm m} =1.0$). In both analyses the mass function is
  convolved with 45 per cent scatter in the overall $M$--$L$
  conversion. For the sake of comparison with the results of B01, we
  used here a BBKS power spectrum with $\Gamma=0.2$ for the shape
  parameter and the mass function by Sheth \& Tormen (1999).}
\label{fi:sigom_rdcs} 
\end{figure}

We compare in Fig.~\ref{fi:sigom_rdcs} the constraints from REFLEX and
RDCS. Quite remarkably such constraints are in good agreement. The
somewhat looser constraints from REFLEX should be understood in terms
of the largely different nature of the two samples, which cover widely
different redshift ranges, and of the different analysis method -- the
RDCS analysis is based on a maximum likelihood approach, which allows
one to extract the whole information provided by the distribution of
clusters in the $L$--$z$ plane. Besides the values of $\sigma_8$, the
agreement between the `geometrical' and `dynamical' measurements
of $\Omega_{\rm m}$ is quite remarkable.  We believe that this finding
represents important support for the reliability of using galaxy
clusters as tracers of the evolution of cosmic structures.

\begin{figure*}
\centerline{ \hbox{
\psfig{file=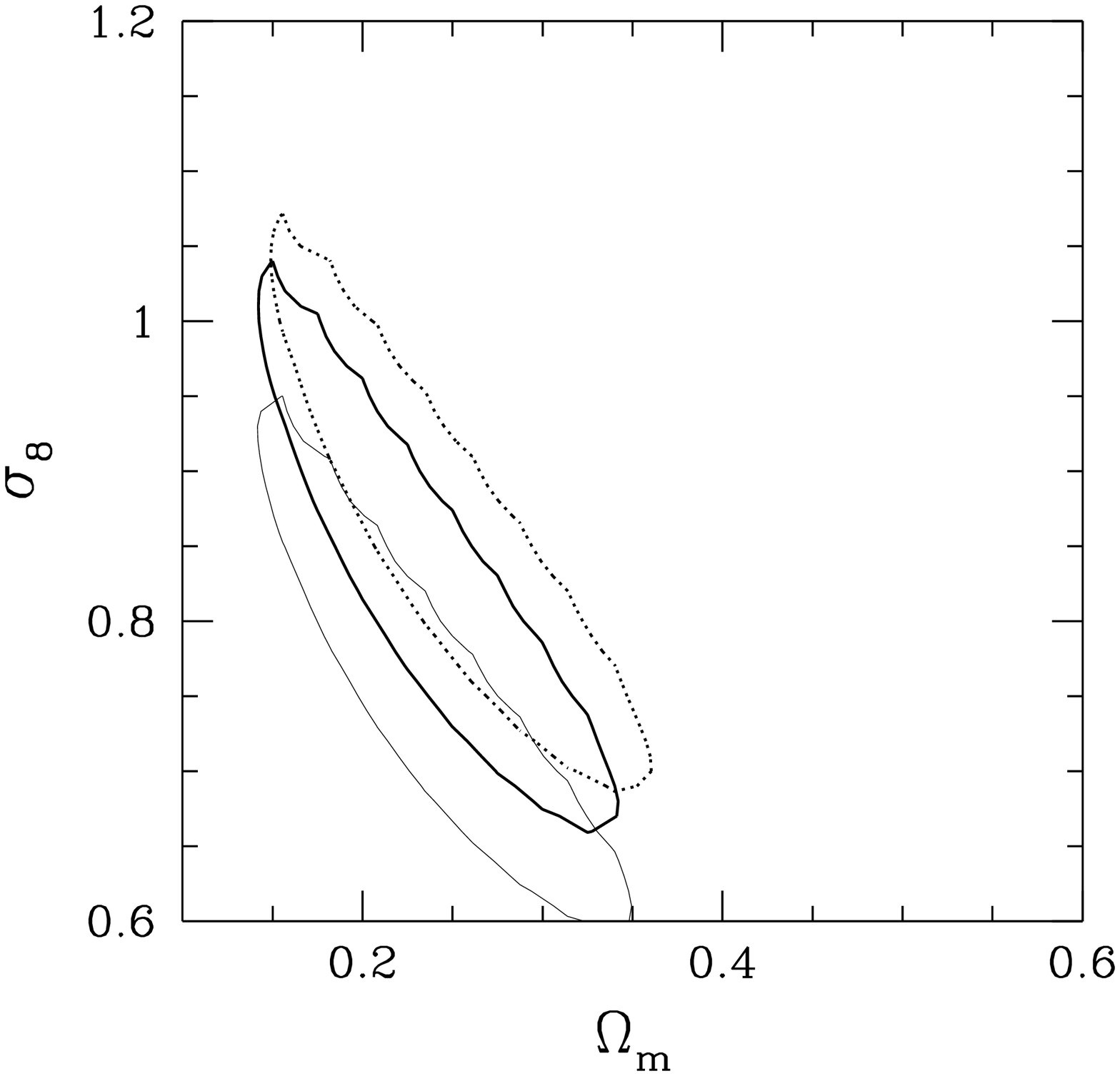,width=6cm} \psfig{file=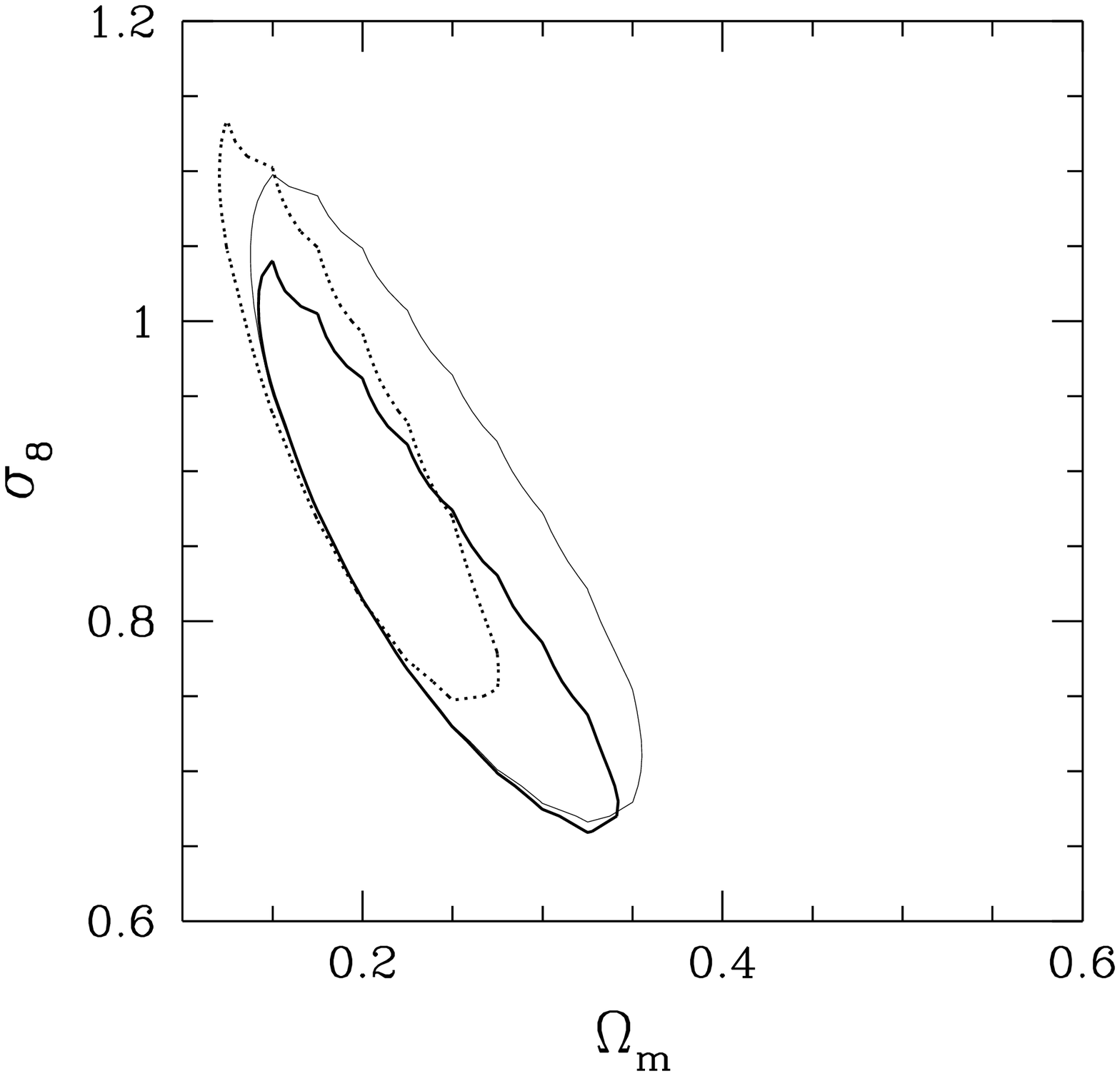,width=6cm}\psfig{file=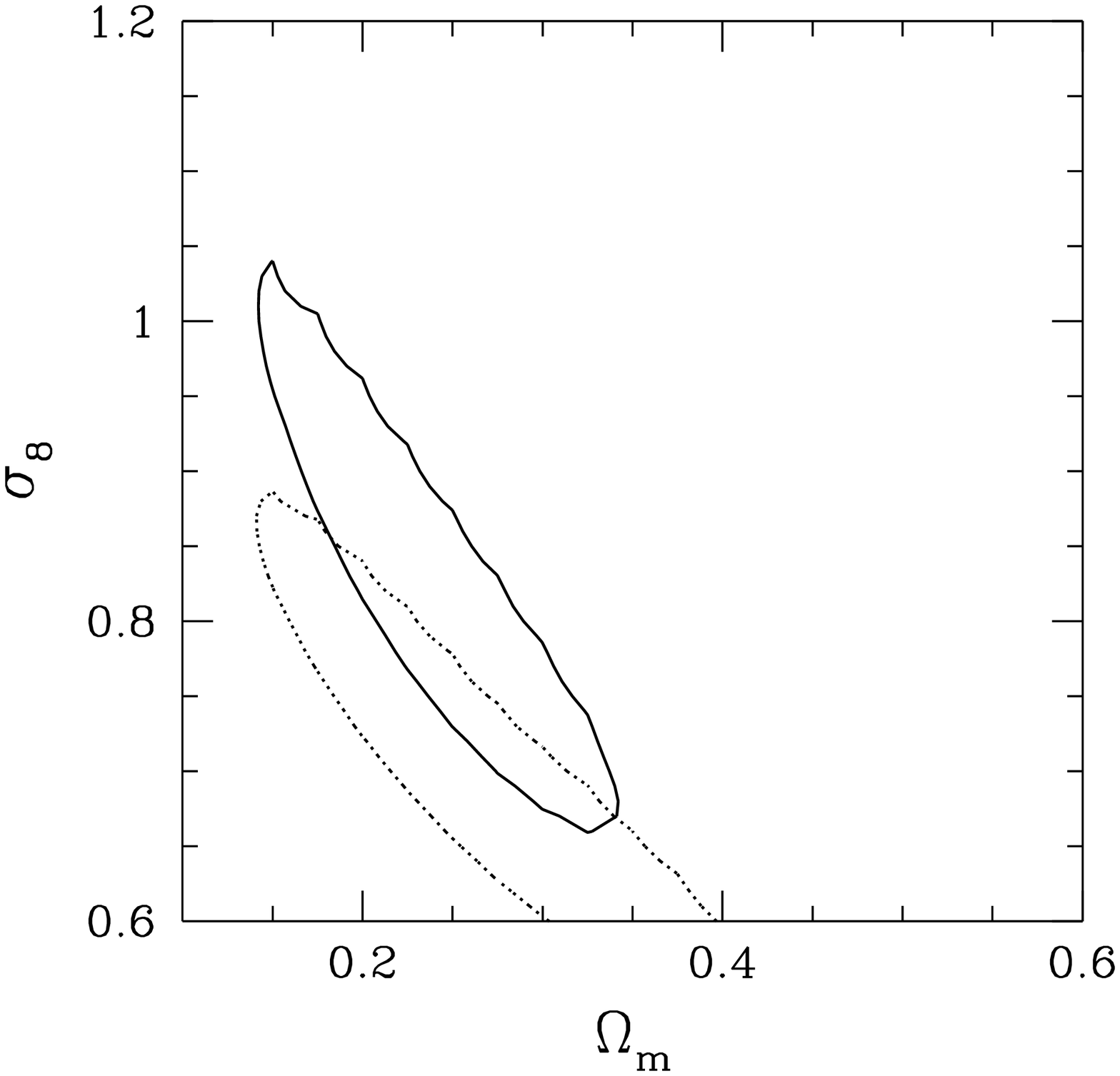,width=6cm}}}
\caption{The 90 per cent confidence contours for the REFLEX XLF, for
  different choices of the $M$--$L$ conversion. In all panels, the
  solid contour corresponds to the reference analysis method, taking
  the best-fitting direct $M_{500}$--$L$ conversion by Reiprich \&
  B\"ohringer (2002, RB02) and convolving the model mass function with
  20 per cent intrinsic scatter. Left panel: the dotted contour is for
  the effect of assuming no intrinsic scatter in the $M_{500}$--$L$
  relation by RB02, while the thin continuous contour is for the
  maximal assumption of 39 per cent intrinsic scatter in the
  $M_{500}$--$L$ relation. Central panel: the thin continuous contour
  shows the effect of marginalising over the amplitude of the $M$--$L$
  relation, which is assumed to vary with a uniform distribution
  of $\pm20$ per cent about the best-fitting value;
  the dotted contour is based on taking
  the inverse $M$--$L$ best-fitting parameters, as provided by RB02,
  in the reference analysis. Right panel: the dotted contour is for
  the $M$--$L$ conversion based on the same method used to obtain the
  solid contour in Fig.7, but with $T_\ast=1.75$ for the $M$--$T$
  relation (long-dashed line in Fig.~\ref{fi:ml}).}
\label{fi:sigom_reip}
\end{figure*}

We show in Fig.~\ref{fi:sigom_reip} how cosmological constraints
alter as we change different aspects of the analysis procedure, such
as the $M$--$L$ conversion, as well as the amount of
uncertainty in this conversion and how it is treated.  Our
reference analysis corresponds to choosing the best-fitting direct
$M$--$L$ relation from RB02, with 20 per cent intrinsic scatter,
convolved with the mass function (results shown with solid contours 
in the panels). In this case, we obtain
\be
\sigma_8\,=\,0.86^{+0.12}_{-0.16}\,;
\Omega_{\rm m}\,=\,0.23^{+0.10}_{-0.06}\,,
\ee
where error bars correspond to 68 per cent uncertainties on the two
parameters. If we fix the density parameter at $\Omega_{\rm m}=0.3$,
we find
\be
\sigma_8\,=\,0.74^{+0.03}_{-0.04}\,,
\ee
where the error corresponds to the 68 per cent uncertainty for one
interesting parameter.
The above quoted errors only reflect the
shot noise, propagated from the Poissonian errors quoted by B02 (and no
systematic uncertainty in the $M$--$L$ relation, which we discuss below).
As pointed out by Evrard et al.~(2002; see also Hu \& Kravtsov 2002 and
White 2002), sample variance due to finite survey size and clustering
should also be considered.  However, given the large size of the
REFLEX survey, we expect sample variance not to be important, compared
with other uncertainties related to the $M$--$L$ conversion.

In the left panel of Fig.~\ref{fi:sigom_reip} we show the effect of
changing the amount of intrinsic scatter in the $M$--$L$ relation. As
expected, assuming a larger scatter decreases the normalization of the
power spectrum. For instance, taking $\Omega_{\rm m}=0.3$ and assuming
zero scatter gives $\sigma_8\simeq 0.8$, while assuming a 39 per cent
scatter lowers $\sigma_8$ to about 0.65.
In the central panel we show the effect of the statistical uncertainty
in the $M$--$L$ relation, by considering both the direct and the
inverse fitting provided by RB02, as well as the effect of
marginalising the likelihood over the $M$--$L$ amplitude.  The effect
of using the inverse fitting (dotted contour) is that of marginally
favouring lower $\Omega_{\rm m}$ values, while keeping the
$\sigma_8$--$\Omega_{\rm m}$ degeneracy direction unchanged. This
indicates that the statistical stability of the $M$--$L$ conversion
has only a small effect. As for the marginalisation procedure (thin
continuous contour), it has been realized by integrating the likelihood
function over a range of $M$--$L$ amplitudes, defined by varying the
best-fitting value by $\pm20$ per cent, with a uniform
distribution. The effect of marginalising is that of widening the
contours, while shifting the most likely $\sigma_8$ to larger values,
since no convolution with the scatter is now performed. In fact, for
$\Omega_{\rm m}=0.3$ we have 
\be 
\sigma_8\,=\,0.79^{+0.06}_{-0.07}\,,
\ee 
for the 68 per cent confidence region.

The right panel shows the effect of using method {\em (a)\/} for the
$M$--$L$ conversion, and assuming $T_\ast=1.75$, as a compromise
between observational and theoretical determinations.  Since the
$M$--$L$ relations from these two methods are quite similar (see
Fig.~\ref{fi:ml}), the slightly higher normalization from method {\em
(b)\/} is precisely the consequence of having assumed a 20 per cent
intrinsic scatter, instead of 45 per cent as in method {\em (a)}.

\begin{figure} 
\centerline{ \hbox{
\psfig{file=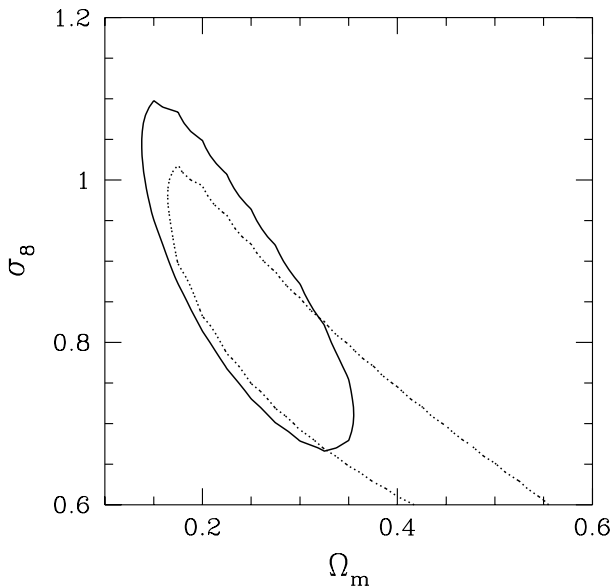,height=8.5cm,width=8.5cm,angle=0} } }
\caption{Comparison between the 90 per cent confidence contours
derived from the REFLEX XLF (continuous line) and the XTF analysis
(dotted line).  The XLF analysis is based on method $(b)$ for the
$M$--$L$ conversion, by marginalising over the amplitude of the
$M$--$L$ relation by RB02, which is varied by $\pm 20$ per cent around
the best fitting value with uniform distribution (see text). The XTF
contour has been obtained by marginalising over $T_\ast$, which is
assumed to vary with uniform distribution from 1.5 to 1.9 (see also
eq.10). Both analyses are based on assuming $\Omega_b=0.019\,h^{-2}$
and $h=0.7$ in the power spectrum by Eisenstein \& Hu (1999).}

\label{fi:xlf_xtf} 
\end{figure}

\section{Discussion}
\subsection{Comparing XTF and XLF results}
We can now compare the XLF and XTF approaches, making as far as is
possible consistent assumptions throughout.  
We show in Fig.~\ref{fi:xlf_xtf} a comparison between the marginalised
constraints derived from the analyses of the previous sections. Both
analyses have been realized by assuming $\Omega_b=0.019\,h^{-2}$ and
$h=0.7$ in the power spectrum by Eisenstein \& Hu (1999). The XLF
constraint is that derived by marginalising over the amplitude of the
$M$--$L$ relation by RB02, on which the result of eq.(15) is based.
The XTF result is derived by marginalising over $T_\ast$, which is
uniformly varied from 1.5 to 1.9. Even bearing in mind the systematics
affecting both the XTF and the XLF analyses, the two methods provide
remarkably consistent constraints. This represents a non-trivial
result, owing to the quite different approaches by which such
constraints have been derived: not only are the cluster samples and
the analysis methods different, but also the systematics associated
with the $M$--$T$--$L$ scalings enter in completely independent ways.
Although different assumptions on how to interpret and treat the
scatter in the $M$--$T$ and $M$--$L$ relations produce different XLF-
and XTF-based constraints, nevertheless it is reassuring that such a
good agreement is just obtained for rather natural choices of these
parameters.  The wider contour from the XTF analysis is due to the
narrower dynamical range probed in cluster masses by our sample of
cluster temperatures, which does not allow to put strong constraints
on the shape of the power spectrum.  In fact, while our XTF samples
about half a decade in temperature, the REFLEX XLF samples about 3.5
decades in cluster luminosity (0.1-2.4 keV band), which correspond to
almost 1.5 decades in temperature from the $L$--$T$ relation of
eq.(3).

The main disagreement between the XTF analysis by PSW and the XLF
analysis by B01 was neither due to the reliability of one of the two
or both samples, nor to the $T_\ast$ assumed (which in fact would have
implied a discrepancy in the opposite direction), but rather to the
procedure of convolving versus marginalising the scatter.  This
illustrates that although the application of the cluster abundance
normalization technique appears to be simple, it involves a chain of
steps, and the source of discrepancies can be rather subtle.  However
it also serves to reinforce our central point -- with sufficient care
the statistical and methodological differences can be reduced to the
point that the major remaining uncertainty is in the normalization of
the mass scale of clusters and in the determination of the nature of
the scatter between the theory--predicted mass and the observed X--ray
luminosity and temperature. 

\subsection{Comparing with other estimates of $\sigma_8$}
\label{sec:estimates}

We list in Table~1 other recent determinations of $\sigma_8$, which
are based on different methods.  Older estimates and associated
references can be found in PSW. Rather than analyse individual studies
in detail, we will make some general remarks which help explain the
differences.

The normalization of density perturbations (through the mass variance
at $8\,h^{-1}$Mpc) can be evaluated by means of several different
observables, either on much larger scales (e.g.~CMB derived
constraints), on scales ${\sim}\,10\,h^{-1}$Mpc (containing the mass
of a rich cluster of galaxies, which is not far into the non-linear
regime), or on much smaller scales (e.g.~through lensing or
Ly$\,\alpha$ forest variance measurements).  Normalization estimates
from large or small scales can be interpreted as $\sigma_8$
determinations through extrapolation of the power spectrum.  This is
cosmology dependent, and hence it should be no surprise that some of
the estimates in Table~1 appear discrepant. Even if we focus on
methods which are directly measuring the mass variance on scales close
to $8\,h^{-1}$Mpc, there still appears to be a bewilderingly wide
range of published values.

\begin{table*}

\begin{center}
\begin{tabular}{l|cccc}
Authors & $\sigma_8$ & error & $\Gamma$ & Method\\
\hline
Van Waerbeke et al.~(2001) & 0.88 & 0.05 & -- & WL \\
Van Waerbeke et al.~(2002) & 0.98 & 0.06 & 0.2  &WL \\
Bacon et al.~(2002) & 0.97 & 0.13 & --  &  WL\\
Refregier et al.~(2002) & 0.93 & 0.17 & 0.21   &WL\\
Hoekstra et al.~(2002) & 0.87 & 0.03 & -- & WL\\
Brown et al.~(2002) & 0.74 & 0.09 & -- & WL\\
Hamana et al.~(2002) & 0.73 & 0.27 & 0.21 &  WL\\ 
Jarvis et al.~(2002) & 0.71 & 0.14 & 0.21     & WL\\

Bahcall et al.~(2002) &  0.72 & 0.06 & -- & OC\\
Viana et al.~(2002) &  0.61 & 0.10 & 0.1   &WLC  \\ 

Blanchard et al.~(2000) & 0.75 & 0.02 & -- & XTF \\
Henry (2000) & 0.77 & 0.15 & -- & XTF \\
Oukbir \& Arnaud (2001) & 0.91 & -- & -- &  XTF \\
Pierpaoli et al.~ (2001)  & 1.02 & 0.07 & 0.23 &  XTF \\ 
Seljak (2001) & 0.77 & 0.06 & 0.20 &  XTF \\

Reiprich \& B{\"o}hringer (2002) & 0.68 & 0.13 & 0.17 &  XLF \\
Borgani et al.~(2001)  & 0.67 & 0.06 & 0.23 &  XLF \\
Schuecker et al.~(2002) & 0.71 & 0.03 & -- &  XLF \\
Allen et al.~(2002) & 0.72 & 0.02 & -- &  XLF \\

Lahav et al.~(2001) & 0.73 -- 0.83 & 0.07 & 0.21 &  PS \\
Szalay et al.~(2001) & 0.91 & 0.06 & 0.19 &  PS \\ 

Bond et al.~(2002) & $\ge 1$ & -- & -- &  SZ PS \\
Komatsu \& Seljak (2002) & 1.05 & 0.05 & -- &   SZ PS \\
\hline
\end{tabular}
\end{center}
\caption{\footnotesize%
Some recent estimates of $\sigma_8$.  The
errors are typically statistical.  We have tried to give values
consistently for $\Omega_{\rm m}\simeq0.3$, while $\Gamma\simeq 0.2$
is a typical value of the shape parameter.  Methods used include: weak
lensing (WL); Optical clusters (OC); clusters normalized using weak
lensing (WLC); X-ray temperature function (XTF); power spectrum (PS);
Sunyaev-Zel'dovich effect power spectrum (SZ). The two errors for the
WL analysis by Refregier et al.~(2002) are for statistical uncertainty
and cosmic variance, respectively.}
\label{tab:estimates}
\end{table*}

One point to realize is that different cosmological assumptions are
made by different sets of authors.  Since the density parameter is
constrained to a first approximation through the combination
$\Omega_{\rm m} h$ (sometimes referred to as the `shape parameter'),
any uncertainty in the value of $h$ translates into a similar
uncertainty when constraining $\Omega_{\rm m}$.  Hence there are
ambiguities here, which one has to be careful about when comparing
results between authors.  Different values of $\sigma_8$ will be
obtained using identical data but either a fixed $h$, fixing
$\Omega_{\rm m}h$, or marginalising over various priors.  This
explains some of the discrepancies in the Table.

There have been several recent results from weak lensing
(Bacon et al.~2002, Brown et al.~2002, Hoekstra et al.~2002,
Jarvis et al.~2002, Refregier et al.~2002, van Waerbeke et al.~2002). 
The quoted $\sigma_8$ values,reported in Table \ref{tab:estimates} in
chronological order, range between 0.7 and 0.9, and have been
decreasing in the past year.  It is early days for this method, since
the formal significance of cosmic shear detection is still rather
modest.  More importantly for the purposes of comparison with cluster
abundance values of $\sigma_8$, most lensing studies have only poor
knowledge of their source redshift distributions and are measuring on
a scale which is a factor of several smaller than the
${\sim}\,10\,h^{-1}$Mpc probed by clusters, and may be affected by
non--linear clustering.
Despite the fact that latest works have done major improvements in
assessing systematic error via the B--mode power spectrum, this still
remains a very hard measurement and the nature of the systematic
errors is not yet well understood.  Estimates using large-scale
structure measurements coupled with Cosmic Microwave anisotropies
(e.g.~Lahav et al.~2001) also give similar values, but are even more
dependent on the adopted cosmology and which parameters are held fixed.

There are many other minor differences between some of the studies
involving the analysis of the cluster number density.  Precisely how
the samples are constructed is obviously a major issue.  For instance,
when dealing with X-ray properties this is particularly an issue for
deciding how best to deal with substructure and cooling flows. Here
different authors make different decisions, and it is to be hoped that
such ambiguities will eventually disappear as the simulations improve
and can be more directly compared with the observations.  Another
issue is merging corrections, which some recent studies have used,
motivated by the idea that Press-Schechter describes the number
density of halos at the virialisation epoch. However (as we stressed
in PSW), it is clear that the effect of merger history is already
implicit in the intrinsic scatter of the $M$--$T$ relation, and so
should not be included twice.  Other minor issues include whether
redshift corrections are made, what profiles are used to convert
between different mass quantities, whether there is curvature in the
adopted $M$--$T$ or $M$--$L$ relations, etc.

To summarize this part: differences in $\sigma_8$ can be due to a
number of factors.  Firstly, the method used may require an
extrapolation to $8\,h^{-1}$Mpc.  Secondly, there are varying
assumptions made about the values of other cosmological parameters in
the analysis.  Thirdly, there are different forms used for the mass
function or the transfer function.  Fourthly, even when the basic
method adopted is the same, there can still be detailed differences in
the approaches of each set of authors (e.g.~which sources of scatter
are included, how the catalogue is constructed or whether a merger
correction is applied).  And finally, for XTF- and XLF-based
approaches there is the choice for the mass--temperature and
mass--luminosity conversion, respectively.

\section{Conclusions} \label{sec:conclusions}

We investigated the determination of $\si_8$ as derived from the XTF
and XLF approaches.  We considered many possible choices in terms of
data and methodology, and paid special attention to the assessment of
possible sources of biases and systematic uncertainties in the
adopted procedure.

In both XTF- and XLF-derived constraints, the most important effect on
$\sigma_8$ estimates is the error in the scaling relation connecting
mass to temperature and to luminosity.  Moreover, different ways of
introducing such an error have a significant impact on the final
$\sigma_8$ range.  If it is introduced as a convolution with the mass
function, its overall effect is to lower the $\sigma_8$ value, while
if it is introduced by marginalising the final likelihood, then the
consequence is mainly to broaden the error contours.  We argue that
the former (latter) procedure is more adequate when dealing with
statistical (systematic) errors. 
For instance, in the marginalisation procedure one has to make a
decision about the functional form of the distribution with which the
marginalised parameters are varied. In our comparison between XTF- and
XLF-based constraints (see Fig. \ref{fi:xlf_xtf}) we decided to assume
a uniform distribution for $T_\ast$ within a finite range. While there
is no rigorous justification for this choice, this procedure allowed
us to have hints about the effect of treating the uncertainties in the
$L$--$T$--$M$ scaling relations in different ways. On the other hand,
in order to treat errors adequately,
one should be able to precisely assess their nature, and at the moment
the split between scatter and overall normalization uncertainties is
not well characterised, nor are these uncertainties precisely
quantified.  We notice, for example, that part of the previous
disagreement between PSW and B01 was not in fact due to the adopted
value of the $M$--$T$ normalization $T_\ast$ (which would have given a
discrepancy in the opposite direction), but rather to the statistical
procedure used.

For the XTF approach, the main scaling relation involved is the
$M$--$T$ one.  Despite the fact that the agreement between
observations and simulations has improved, mainly due to better
modeling of the cluster physics in the simulations (e.g.~Muanwong et
al.~2002; Dav\'e et al.~2002; Tornatore et al.~2002, in preparation),
there is still a significant scatter in the quoted $T_\ast$
values. Should that be taken as the `systematic' error, it would
certainly dominate the statistical error quoted by either observers or
simulators (which is typically about 10 per cent).  At the same time,
observational mass determinations are probably affected by significant
biases, due for instance to the necessary extrapolation from the
(small) internal region observed to the larger one involved in the
mass function.  Other source include possible deviations of the
surface brightness profile from simplified fitting models and to the
violation of the often made hydrostatic and isothermal assumptions.

The same general comments apply to the XLF approach.  In this case,
the fairly large number of clusters included in nearby and distant
samples means statistical noise is unimportant for the resulting
uncertainties. In fact the same complex physical processes which
affect the ICM temperature are also very important when dealing with
X-ray luminosity, which is highly sensitive to the local properties of
the ICM. Therefore, more robust constraints on $\sigma_8$ will require
substantial efforts to improve both the observational picture and the
theoretical understanding of the relevant physical processes, which
determine the X-ray properties of the intra-cluster medium.

Because of the variety of $\sigma_8$ values in the literature, we
carried out careful tests of the impact of different assumptions on
the final result.  On the theoretical side, it is important to rely on
the most accurate mass function for the model being considered.  In
our analysis, we have used the Jenkins et al.~(2001) mass function,
and found that it gives closer results to pure Press--Schechter than
to Sheth--Tormen for the mass range probed.  Moreover, the use of the
explicit $\sigma_8$ quantity instead of fitting formulae, as well as
an accurate transfer function which properly includes baryonic
suppression, is also necessary in order to reach the desired accuracy.
While such effects are probably small with respect to the systematic
uncertainties in the $M$--$T$--$L$ relations, nevertheless they should
be properly understood if the cluster abundance normalization is to be
determined with high precision.

Bearing in mind all these caveats, a few firm conclusions can be drawn
from our analysis.  Firstly, current data on the XLF and XTF for
nearby clusters give quite consistent results on cosmological
parameters, with $\sigma_8 \simeq 0.8$ for a flat $\Omega_{\rm m}=0.3$
Universe. While one would be tempted to attach a $\sim 5$ per cent
statistical error (at $1\sigma$) to this determination, it is
fair to say that at least twice as large an error, associated with
uncertainties in the $M$--$T$ and, especially, in the $M$--$L$
relations, is more realistic.
Secondly, under the same assumptions for the $M$--$L$ conversion in
the XLF analysis, dynamical constraints on the $(\Omega_{\rm
M},\sigma_8)$ plane, derived from the evolution of the cluster
population (e.g.~B01), are fully consistent with the geometrical
constraints derived from nearby samples.

The cluster abundance as a cosmological probe has been developing very
fast
in the past few years.
New observations have allowed an empirical determination of the scaling 
relations involved, while more refined simulations are helping 
to reconcile past discrepancies with observations. 
Furthermore, great effort has been devoted in understanding 
which modelling should be used in terms of transfer and
mass function, and to the statistical procedures to compare observed and
predicted number density of galaxy clusters. Furher improvements are
expected also 
on the observational side, thanks to the improved capability of X-ray
satellites of the last generation to precisely measure cluster
temperatures.
A cereful study over a statistically significant sample of cluster will
therefore allow a better calibration of the scaling relations and of
their intrinsic scatter. 
At the same time, improvements in simulations, in terms of both
resolution and physics included, will allow a reliable framework for the
interpretation of the observational results.

X-ray surveys, both in contiguous areas and based on cluster searches
from {\sl Chandra\/} and {\sl XMM\/} archives, will certainly help in
improving the quantity and quality of cluster data.
Future surveys, using missions like {\sl WFXT\/} (Burrows et al.~1992)
or {\sl DUET}, as well as Sunyaev-Zel'dovich surveys (see Table 1 of
Schulz \& White 2002 for a list), will be extremely valuable for increasing
the statistics of clusters at $z>1$ by orders of magnitude, to enable
evolutionary studies to be carried out.
However, what is really needed is a robust set of mass estimates for a
sample of clusters.
An inter-comparison of multi-wavelength mass estimates, from X-ray data,
velocity dispersions, gravitational lensing and Sunyaev-Zel'dovich
measurements should enable systematic biases to be much better
understood than they are at the moment.  On the theoretical side, the
challenge for the future will be to realize a new generation of
cluster simulations, where the inclusion of the relevant physical
processes should be coupled with a large enough dynamical range so as
to accurately resolve the internal structure of the clusters while
encompassing cosmological volumes.
It will require a concerted effort from both the theoretical and observational
communities, tackling the many remaining obstacles, in order to hone the
cluster abundance into a precision cosmological tool and to keep it
competitive with rapid advances in the other methods.

\section*{Acknowledgments}
EP is supported by NASA grant NAG5-11489, 
DS by NSERC--Canada and
MW by NSF, NASA, DoE and a Sloan Foundation Fellowship.
The authors wish to
thank Hans B\"ohringer, Luigi Guzzo, George Ellis, Vincent Eke, Piero
Rosati and Peter Schuecker for helpful and stimulating discussions.
EP and MW are grateful to the Aspen Center for Physics and to the
KITP for hospitality during the
preparation of this work;  EP is also grateful to the `Dipartimento di
Astronomia di Trieste'  and DS is grateful to the Institute for 
Astronomy, Edinburgh.

\end{document}